\documentclass{article}
\usepackage{spconf,amsmath,graphicx}

\usepackage{makecell}
\usepackage{xcolor}
\usepackage{url}
\usepackage{multirow}
\usepackage{subcaption}
\usepackage[T1]{fontenc}
\usepackage{booktabs}

\usepackage{pgfplots}

\title{Learning in your voice: Non-parallel voice conversion based on speaker consistency loss}
\name{Yoohwan Kwon$^{1,2}$, Soo-Whan Chung$^{1}$, Hee-Soo Heo$^{2}$, Hong-Goo Kang$^{1}$}

\address{
$^{1}$Department of Electrical \& Electronic Engineering, Yonsei University, South Korea\\
$^{2}$Naver Corporation, South Korea
}

\begin{document}
%
\maketitle
\begin{abstract}
In this paper, we propose a novel voice conversion strategy to resolve the mismatch between the training and conversion scenarios when parallel speech corpus is unavailable for training.
Based on auto-encoder and disentanglement frameworks, we design the proposed model to extract identity and content representations while reconstructing the input speech signal itself. 
Since we use other speaker's identity information in the training process, the training philosophy is naturally matched with the objective of voice conversion process.
In addition, we effectively design the disentanglement framework to reliably preserve linguistic information and to enhance the quality of converted speech signals.
The superiority of the proposed method is shown in subjective listening tests as well as objective measures.
\end{abstract}

\section{Introduction}
\label{sec:intro}
Voice conversion~(VC) is a technique for converting the speech of source speaker to the speech that sounds like being spoken by a target speaker while preserving the linguistic information of source speech.
In early studies~\cite{mohammadi2014voice,chen2014voice}, most voice conversion researches were performed using parallel speech corpus, i.e. speech signals with common scripts spoken by many different speakers.
These methods effectively transfer ones' speaking style to others while preserving speaking contents. However, they result in severely distorted sound if the test scripts to be converted are not included in the training stage.
Although the problem can be relieved by using large-scale datasets, it requires a large amount of costs for manual annotations and parallel speech recording.

On the Internet, there are plethora of utterances spoken by a large number of people, and some large-scale datasets are publicly available.
In recent, these non-parallel speech corpus is used for voice conversion researches that are implemented independent of speaking contents.
It is a highly challenging task since there is no explicit target signal in supervision.
Instead of relying on strong supervision, some works use additional data or extra systems helpful to the conversion process such as text scripts~\cite{zhang2019non} or speech recognition~\cite{xie2016kl,saito2018non}.
These approaches are somewhat effective when we can not use strong supervision, but it is still not a proper solution.
To solve the problem, disentanglement~\cite{wu2020vqvc+,wu2020one} and 
generative model-based approaches~\cite{hsu2017voice,kameoka2018acvae} were studied.
Especially, generative adversarial network~(GAN)-based methods in ~\cite{kaneko2018cyclegan,kaneko2019cyclegan,kameoka2018stargan,kaneko2019stargan} showed promising results in the conversion tasks.
However, these methods are still suffered from low speech quality with low naturalness and inaccurate pronunciation on outputs. 

AutoVC~\cite{autovc} is the most prevalent work that are related to the proposed method, where it uses auto-encoder framework to generate speech signals.
In the training phase, AutoVC decomposes input signals into identity and content features and reconstructs itself to build a speech generation system.
In the conversion phase, it replaces identity feature with other's representation to convert speech into target's voice.
AutoVC is classified into a zero-shot learning approach since it does not require any additional cues or systems and is applicable to unseen speakers.
However, the training criterion of AutoVC is set to reconstruct input signal itself, not to convert speaking styles or maintain contextual information; thus, the converted speech has distortions or missing pronunciation.

In this paper, we present a novel training strategy, whose objective is well-matched with voice conversion.
The trunk structure is the same as the one used in AutoVC, i.e. reconstructing input spectral features using disentangled embeddings.
In addition to the self-reconstruction process, we replace the input identity embeddings with the ones taken from target speakers.
Consequently, the self-reconstruction helps to learn speech generation, and the additional speaker consistency criterion provides a supervision for speaking style conversion.
Additionally, we modify the disentanglement method for content embeddings to preserve linguistic expression on converted speech.
To preserve reliable linguistic expression on converted speech, we also introduce an attention method to encode content embeddings instead of the conventional hard sampling strategy.
Our contributions to the conversion task improve the voice conversion performance by addressing the conversion process on the training stage and introducing a powerful disentanglement method.
The superiority of the proposed method would be examined in several objective and subjective measurements.

\begin{figure*}[t!]
\centering
     \begin{subfigure}[b]{0.45\linewidth}
         \centering
         \includegraphics[width=\textwidth]{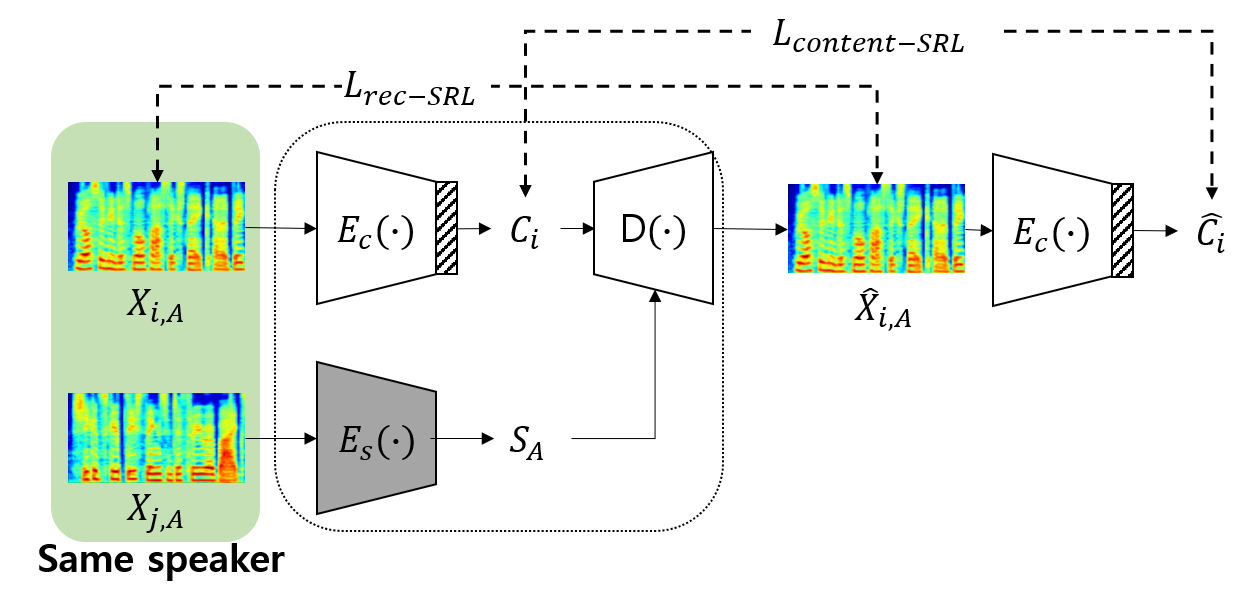}
         \caption{Training strategy with a self-reconstruction loss}
         \label{fig:pos}
     \end{subfigure} 
     \begin{subfigure}[b]{0.45\linewidth}
         \centering
         \includegraphics[width=\textwidth]{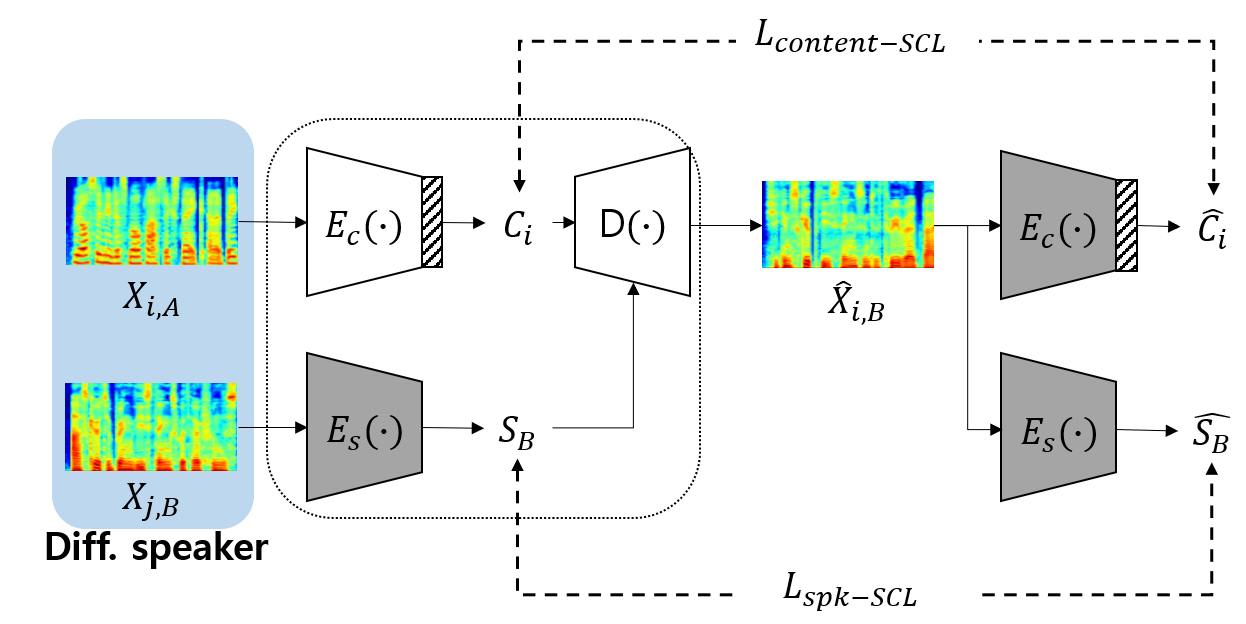}
         \caption{Training strategy with a speaker consistency loss}
         \label{fig:neg}
     \end{subfigure}    
     \caption{Overview of the proposed learning strategy. Gray-coloured blocks are pretrained, not updated in the training phase.}
     \label{overview}
     \vspace{-5pt}
\end{figure*}

\section{Related Works}
\label{sec:related_work}

\subsection{AutoVC}
\vspace{-3pt}
AutoVC is a zero-shot non-parallel many-to-many voice conversion model based on a simple auto-encoder framework~\cite{autovc}.
By decomposing speech inputs into content and identity representations with information-constraining bottleneck encoding layers, the model reconstructs the input speech in the decoding stage.
The content encoder takes input speech and passes it through a bottleneck to produce a content embedding conditioned on source speaker embedding from a pretrained identity encoder.
The content embeddings are downsampled to make it more difficult to present speaker characteristics of input signals. 
Then the decoder takes the content embedding and identity embedding to reconstruct spectral features.
When training, source speaker embedding is fed to the decode for self-reconstruction supervision.
The network is trained to minimize reconstructed features and input features, and the content embedding is extracted from the reconstructed features to make it resemble that of input speech by minimizing the distance between them.
Although AutoVC shows powerful disentanglement and conversion performance, there are still rooms for improvement.

\vspace{-5pt}
\subsection{Style transfer with perceptual loss}
\vspace{-3pt}
On various generation tasks such as style transfer~\cite{johnson2016perceptual,kotovenko2019content}, voice conversion~\cite{kameoka2018acvae}, there is commonly a lack of ground truth data.
Perceptual loss has been widely used to represent target information on the generation outputs, and it is advantageous on learning tasks using non-parallel data.
Perceptual loss does not directly address the objective of tasks, e.g., minimizing differences between pixels or spectral features, 
but it regulates representations of outputs on the latent space.
It minimizes the distance between high-level representations of outputs and that is extracted from other data presenting the desired feature.
This learning strategy is powerful to provide supervision of perceptual and semantic expressions on outputs while not forcing outputs to match target data directly.
\vspace{-5pt}

\section{Proposed method}
\label{sec:proposed}
In this section, we demonstrate the proposed learning strategy for voice conversion using a disentanglement framework.
The common problem with using a non-parallel dataset on voice conversion is that there is no explicit ground-truth speech to generate.
Our proposed method adopts the strategy of AutoVC and designs a consistency loss to resolve this problem.

\vspace{-3pt}
\subsection{Self-reconstruction loss}
\label{ssec:propose-1}
\vspace{-3pt}
As described in~\cite{autovc}, they set up an auto-encoder to generate input speech signal itself.
Content embedding is extracted from input signal, and identity embedding is brought from another utterance spoken by the same speaker.
Since both speech signals are spoken by the same person, the identity representation should be same.
The decoder takes these content and identity embeddings to reconstruct input spectral features.
Figure~\ref{overview}(a) shows our training strategy using the self-reconstruction loss.
In the self-reconstruction, the objective function is given as follows,
\begin{equation}
    \label{eq:recon}
    L_{SRL} = \parallel D(C_{i,A}, S_{j,A}) - X_{i,A}\parallel_{2} + |{\hat{C}_{i}-C_{i}}|,
\end{equation}
where $D$ is the decoder network, $C_{i,A}$ is content embedding extracted from $i$-th speech $x_{i,A}$ spoken by speaker $A$, and $S_{j,A}$ is the identity embedding of the $j$-th utterance spoken by speaker $A$, $X_{i,A}$ is the spectral feature of $x_{i,A}$, respectively.
\vspace{-3pt}

\subsection{Speaker consistency loss}
\label{ssec:propose-2}
\vspace{-3pt}
In a machine learning-based approach, matching the test and training scenario always achieves better performance than an unmatched scenario.
Since the learning criterion of AutoVC does not directly address the conversion task, we propose a new training strategy to learn spectral modeling in others' voices for effective voice conversion, which is called a speaker consistency loss (SCL).
Our method utilizes an implicit representation as a target for supervision instead of target spectral features, which does not exist in a non-parallel dataset.
The content and speaker embeddings for the reconstruction should be kept on the converted speech signals.
Thus, if we extract content and identity embeddings from converted signals, they should be represented in the same distributions of input embeddings.
By using these characteristics of voice conversion, we exploit other's speech signals for training.
Figure~\ref{overview}(b) shows our training strategy of the speaker consistency loss.
The learning criterion is to minimize distances between embeddings as below:
\begin{equation}
    L_{SCL} = |{\hat{C}_{i}-C_{i}}|+|{\hat{S}_{B}-S_{B}}|,
\end{equation}
where $S_B$ indicates speaker embedding of speaker $B$. 
Here, the content encoder extracts linguistic representations, but its parameters are not updated with the consistency loss.
Whole network is trained with below objective function:
\begin{equation}
    L_{total} = L_{SRL} +\lambda L_{SCL}
    \label{total_loss}
\end{equation}
We use 0.5 for the value $\lambda$, and the combination of multiple learning criteria is beneficial to its training by addressing both voice conversion and speech generation.

\vspace{-3pt}
\subsection{Attentive bottleneck}
\label{ssec:propose-3}
\vspace{-3pt}
\begin{figure}[t]
    \centering
    \includegraphics[width=0.7\linewidth]{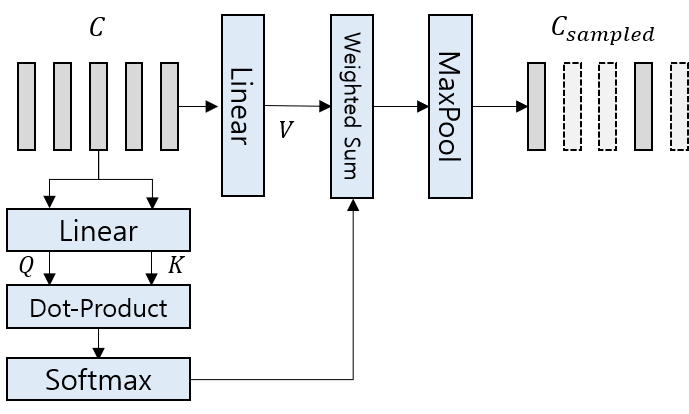}
    \caption{Structure of modified attention bottleneck block}
    \label{fig:bottleneck}
\vspace{-10pt}
\end{figure}
The other key factor of voice conversion is feature disentanglement.
AutoVC disentangles content embeddings by an information bottleneck structure with temporal downsampling as described in~\cite{autovc}.
However, due to its structural limitation of hard sampling, it frequently loses content information.
Therefore, we modify the information bottleneck structure using attention mechanism.
We exploit a self-attention structure to extract content embedding as in Figure~\ref{fig:bottleneck}.
The enhanced content feature that contains linguistic information of near frames compensates the loss from temporal downsampling.
Scaled dot-product~\cite{vaswani2017attention} is used for attention mechanism.
After that, we change the downsampling to a max-pooling procedure.
Since max-pooling does not take a value from the fixed frames unlike downsampling, the attention mechanism works well to contain the linguistic information while maintaining its disentanglement effect of downsampling.
\section{Experiments}
\label{sec:experiments}
In this section, we conduct several objective and subjective experiments on a non-parallel many-to-many voice conversion task to confirm the performance of the proposed method.

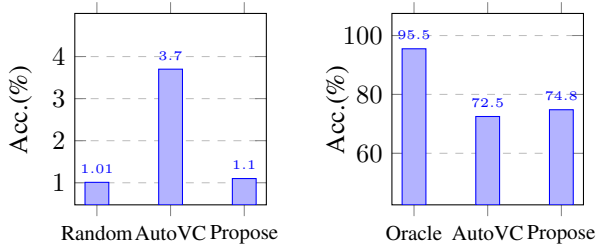
\begin{figure}[t]
\begin{minipage}[t]{0.48\columnwidth}
    \begin{tikzpicture}
            \begin{axis}
            [
                ybar,
                enlargelimits=0.15,
                legend style={at={(0.5,-0.4)},
                  anchor=north,legend columns=-1},
                ylabel={Acc.(\%)},
        	    y label style={at={(0.2,0.5)}},
                symbolic x coords={Random, AutoVC, Propose},
                x tick label style={font=\footnotesize},
                xtick=data,
                nodes near coords,
                node near coords style={font=\tiny},
                width=\columnwidth, height=\columnwidth,
                ymin=1.0, ymax=4.5,
                ytick={1,2,3,4},
                bar width=9pt,
                ymajorgrids=true,
                grid style=dashed
                ]
            \addplot coordinates {(Random,1.01)  (AutoVC,3.7) (Propose,1.1)}; 

            \end{axis}
        \end{tikzpicture}
\end{minipage}
\medskip
\begin{minipage}[t]{0.48\columnwidth}
     \begin{tikzpicture}
            \begin{axis}
            [
                ybar,
                enlargelimits=0.15,
                legend style={at={(0.5,-0.4)},
                  anchor=north,legend columns=-1},
                ylabel={Acc.(\%)},
    	        y label style={at={(0.2,0.5)}},
                symbolic x coords={Oracle, AutoVC, Propose},
                x tick label style={font=\footnotesize},
                xtick=data,
                nodes near coords,
                node near coords style={font=\tiny},
                width=\columnwidth, height=\columnwidth,
                ymin=50.0, ymax=100.0,
                ytick={60,80,100},
                bar width=9pt,
                ymajorgrids=true,
                grid style=dashed
                ]
            \addplot coordinates {(Oracle,95.5) (AutoVC,72.5) (Propose,74.8)}; 
            \end{axis}
        \end{tikzpicture}
\end{minipage}
\caption{Speaker recognition accuracy. \textbf{Left}: the recognition performance of content bottleneck feature. \textbf{Right}: the recognition accuracy on voice conversion outputs}
\label{acc}
\end{figure}

\subsection{Experimental settings}
\label{ssec:settings}
\vspace{-3pt}
We split this dataset into training and test sets; in the evaluation set, we reserve 10 speakers not included in the training set.
Also, we randomly select 10\% of utterances spoken by 99 speakers for the evaluation of seen speaker conversion.
The input speech signal is transformed into an 80-dimensional Mel-spectrum in the logarithm scale, which is sliced at every 16ms with 64ms window size.

The network structure is exactly same as the one in AutoVC~\cite{autovc}, and we use WaveNet vocoder~\cite{oord2016wavenet} to generate speech signal from reconstructed Mel-spectrum.
The speaker encoder consists of 2 LSTM layers, and it is pre-trained in the generalized end-to-end loss~\cite{wan2018generalized} but is not fine-tuned with a training process related to voice conversion.
We conduct several baseline methods proposed in~\cite{autovc} and~\cite{chou2019one} in scratch.

\subsection{Objective evaluation}
\vspace{-3pt}
\label{ssec:disentangle}
We evaluate the performances of the disentanglement and voice conversion strategy for objective comparisons, respectively.
In the evaluation of disentanglement, content bottleneck features are examined in speakers' identities, where they should not specify the speaker if they are well-disentangled from the speaker's identities.
The speaker of converted speech should be recognized as the target speaker, not the speaker of the input speech.

For the speaker recognition on the bottleneck feature, we train an additional back-end classifier from the bottleneck feature, using speakers in the training set.
Figure~\ref{acc} shows the classification accuracy of seen speakers in the evaluation set.
If there remains nothing related to identity on the content bottleneck, it would show 1\% accuracy, whereas our proposed method shows 1.1\%, and AutoVC results in 3.7\%.
The speaker recognition accuracy on the input speech shows 95.5\%, while the results of conversion outputs are 72.5\% for the proposed method and 74.8\% for AutoVC baseline.
These recognition results prove that the proposed method outperforms the baseline method in the conversion strategy as well as the improved disentanglement method.
\vspace{-5pt}
\subsection{Subjective evaluation}
\label{ssec:sub_eval}
\vspace{-3pt}
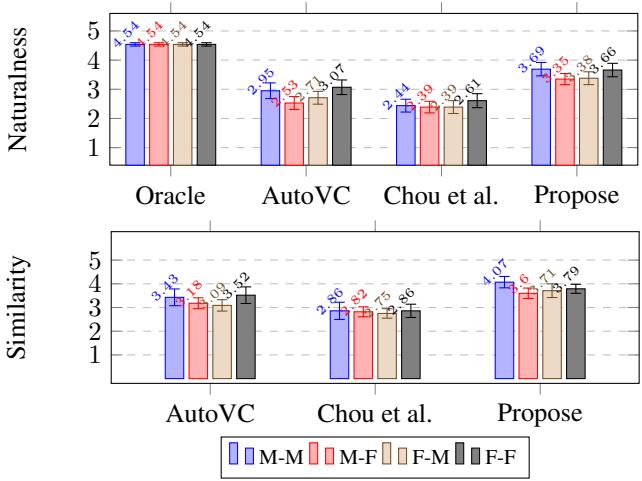
\begin{figure}[t]
\centering
    \begin{minipage}[t]{\linewidth}
    \centering
        \begin{tikzpicture}
            \begin{axis}
            [
                ybar,
                enlargelimits=0.15,
                legend style={at={(0.5,-0.4)},
                  anchor=north,legend columns=-1,font=\fontsize{8}{8}\selectfont},
                ylabel={Naturalness},
                symbolic x coords={Oracle, AutoVC,Chou et al.,Propose},
                xtick=data,
                nodes near coords,
                node near coords style={font=\tiny, rotate=45},
                width=\linewidth, height=3.6cm,
                ymin=1.0, ymax=5.0,
                ytick={1,2,3,4,5},
                bar width=7pt,
                ymajorgrids=true,
                grid style=dashed
                ]
            \addplot+ [
                        error bars/.cd,
                        y dir=both,
                        y explicit,
                    ] 
                    coordinates {(Oracle,4.54) +- (0,0.057)
                                  (Propose,3.69) +- (0,0.23)
            		               (AutoVC,2.95) +- (0,0.27)
            		               (Chou et al.,2.44) +- (0,0.22)
            		               }; 
            \addplot+ [
                        error bars/.cd,
                        y dir=both,
                        y explicit,
                    ] 
                    coordinates {(Oracle,4.54) +- (0,0.057)
                                  (Propose,3.35) +- (0,0.19)
            		               (AutoVC,2.53) +- (0,0.22)
            		               (Chou et al.,2.39) +- (0,0.20)
            		               }; 	
            \addplot+ [
                        error bars/.cd,
                        y dir=both,
                        y explicit,
                    ] 
                    coordinates {(Oracle,4.54) +- (0,0.057)
                                  (Propose,3.38) +- (0,0.22)
            		               (AutoVC,2.71) +- (0,0.22)
            		               (Chou et al.,2.39) +- (0,0.22)
            		               };        
            \addplot+ [
                        error bars/.cd,
                        y dir=both,
                        y explicit,
                    ] 
                    coordinates {(Oracle,4.54) +- (0,0.057)
                                  (Propose,3.66) +- (0,0.23)
            		               (AutoVC,3.07) +- (0,0.25)
            		               (Chou et al.,2.61) +- (0,0.24)
            		               };           		               
            \end{axis}
        \end{tikzpicture}
    \end{minipage}
    \begin{minipage}[b]{\linewidth}
        \centering
        \begin{tikzpicture}
            \begin{axis}
            [
                ybar,
                enlargelimits=0.3,
                legend style={at={(0.5,-0.35)},
                  anchor=north,legend columns=-1,font=\fontsize{8}{8}\selectfont},
                ylabel={Similarity},
                symbolic x coords={AutoVC,Chou et al.,Propose},
                xtick=data,
                nodes near coords,
                node near coords style={font=\tiny, rotate=45},
                width=\linewidth, height=3.6cm,
                ymin=1.0, ymax=5.0,
                ytick={1,2,3,4,5},
                bar width=7pt,
                ymajorgrids=true,
                grid style=dashed
                ]
            \addplot+ [
                        error bars/.cd,
                        y dir=both,
                        y explicit,
                    ] 
                    coordinates {
                                  (Propose,4.07) +- (0,0.24)
            		               (AutoVC,3.43) +- (0,0.353)
            		               (Chou et al.,2.86) +- (0,0.363)
            		               }; 
            \addplot+ [
                        error bars/.cd,
                        y dir=both,
                        y explicit,
                    ] 
                    coordinates {
                                  (Propose,3.60) +- (0,0.22)
            		               (AutoVC,3.18) +- (0,0.23)
            		               (Chou et al.,2.82) +- (0,0.21)
            		               }; 	
            \addplot+ [
                        error bars/.cd,
                        y dir=both,
                        y explicit,
                    ] 
                    coordinates {
                                  (Propose,3.71) +- (0,0.29)
            		               (AutoVC,3.09) +- (0,0.24)
            		               (Chou et al.,2.75) +- (0,0.2)
            		               };        
            \addplot+ [
                        error bars/.cd,
                        y dir=both,
                        y explicit,
                    ] 
                    coordinates {
                                  (Propose,3.79) +- (0,0.19)
            		               (AutoVC,3.52) +- (0,0.35)
            		               (Chou et al.,2.86) +- (0,0.28)
            		               };           		               
            \legend{M-M,M-F,F-M,F-F}
            \end{axis}
        \end{tikzpicture}
    \end{minipage}
\caption{MOS test scores on seen speaker set}
\label{mos-seen}
\vspace{-11pt}
\end{figure}
We conduct three subjective assessments for the quality of speech and its speaker similarity to the target speech; mean-opinion-score (MOS) on naturalness and similarity, and preference tests.
We reserve two types of speaker sets, which are seen and unseen speakers to compare the effects on the conversion of not-enrolled speakers.
Each dataset has another 4 gender pair source-target subsets, i.e., male-to-male (M-M), male-to-female (M-F), female-to-male (F-M), and female-to-female (F-F).
We randomly generated 4 conversion samples for each subset, and 15 listeners were asked to rate the performance of a total of 32 samples per each method.
In the MOS tests, listeners score the naturalness and the similarity of conversion outputs from 1 to 5.
Besides, they compare the conversion methods by judging which sample is more similar to the target speaker for the preference test.

In Figure~\ref{mos-seen} and Figure~\ref{mos-unseen}, it reports the MOS results on the seen and unseen speaker evaluation sets.
Baseline methods in \cite{autovc,chou2019one} show 2.81 and 3.30 MOS on naturalness respectively, while they score 2.62 and 3.09 on similarity.
On the other hands, the proposed method shows significantly improved performance, scoring 3.52 on naturalness and 3.78 on similarity.
In the unseen speaker set, the proposed method is also effective compared to the baselines, where the proposed method shows 3.20 on the naturalness and 3.64 on similarity while AutoVC scores 2.62 and 3.09, respectively.

\begin{figure}[t]
\centering

    \begin{minipage}[b]{\linewidth}
        \begin{tikzpicture}
        \begin{axis}[
            xbar stacked,
        	bar width=7.5pt,
        	nodes near coords,
        	point meta=rawx,
        	node near coords style={font=\tiny},
            legend style={at={(0.5,1.4)},
            anchor=north,legend columns=-1,font=\fontsize{8}{8}\selectfont},
            xmin=0,
            xmax=100,
            symbolic y coords={F-F, F-M, M-F, M-M},
            ytick=data,
            height=3.25cm,
            width=\linewidth,
            enlarge y limits={abs=8pt},
            ]
         \addplot+[xbar,node near coord style={xshift=-50pt,,anchor=center}] plot coordinates {(60.71,M-M) (57.14,M-F) (66.07,F-M) (58.93,F-F)};
         \addplot+[xbar,node near coord style={xshift=-30pt,,anchor=center}] plot coordinates {(37.5,M-M) (33.93,M-F) (28.57,F-M) (26.79,F-F)};
         \addplot+[xbar,node near coord style={xshift=-10pt,,anchor=center}] plot coordinates {(1.79,M-M) (8.93,M-F) (5.36,F-M) (14.28,F-F)};

        \end{axis}
        \end{tikzpicture}
    \end{minipage}
    \begin{minipage}[b]{\linewidth}
        \begin{tikzpicture}
        \begin{axis}[
            xbar stacked,
        	bar width=7.5pt,
        	nodes near coords,
        	point meta=rawx,
        	node near coords style={font=\tiny},
            legend style={at={(0.5,-0.30)},
              anchor=north,legend columns=-1, font=\fontsize{8}{8}\selectfont},
            xmin=0,
            xmax=100,
            symbolic y coords={F-F, F-M, M-F, M-M},
            ytick=data,
            height=3.25cm,
            width=\linewidth,
            enlarge y limits={abs=8pt},
            ]
         \addplot+[xbar,node near coord style={xshift=-50pt,,anchor=center}] plot coordinates {(42.86,M-M) (78.57,M-F) (73.81,F-M) (42.86,F-F)};
         \addplot+[xbar,node near coord style={xshift=-25pt,,anchor=center}] plot coordinates {(45.24,M-M) (16.67,M-F) (23.81,F-M) (40.48,F-F)};
         \addplot+[xbar,node near coord style={xshift=-10pt,,anchor=center}] plot coordinates {(11.9,M-M) (4.76,M-F) (2.38,F-M) (16.66,F-F)};
    
        \legend{\strut \textbf{Proposed}, \strut Neutral, \strut AutoVC}
        \end{axis}
        \end{tikzpicture}
    \end{minipage}
    \caption{Preference results on speaker similarity. 'Neutral' means that they are similar perceptually. \textbf{Upper}: results on seen speaker, \textbf{Bottom}: results on seen speaker}
    \label{abx}
\vspace{-5pt}
\end{figure}
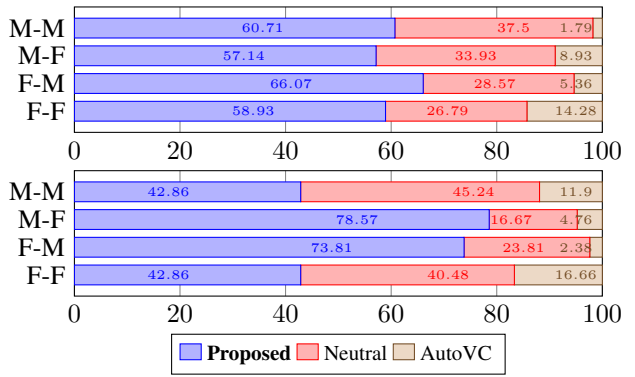

On the preference test in Figure~\ref{abx}, the proposed method is more preferred than AutoVC for entire evaluation sets.
Listeners responded that the outputs of the proposed method are better than those of AutoVC with 60.7\% and 59.5\% preference scores on seen and unseen speaker sets.
However, they reported that if speakers are converted between the same genders, it is frequently difficult to tell the difference between the proposed method and the baseline, although the proposed one is still better than the baseline.
The most impressive results on the preference test is that conversion between different genders with the proposed method is far preferred, while there is little preference on the baseline method.
The conversion samples are available on the demo page\footnote{\url{http://github.com/yoohwankwon/Learning_in_your_voice}}.
\vspace{-5pt}

\begin{figure}[t]
\centering

    \begin{minipage}[b]{\linewidth}
    \centering
    \begin{tikzpicture}
        \begin{axis}
        [
            ybar,
            enlargelimits=0.15,
            legend style={at={(0.5,-0.3)},
            anchor=north,legend columns=-1,font=\fontsize{8}{8}\selectfont},
            ylabel={Naturalness},
            symbolic x coords={Oracle,AutoVC,Chou et al.,Propose},
            xtick=data,
            nodes near coords,
            node near coords style={font=\tiny, rotate=45},
            width=\linewidth, height=3.6cm,
            ymin=1, ymax=5,
            ytick={1,2,3,4,5},
            bar width=7pt,
            ymajorgrids=true,
            grid style=dashed
            ]
        \addplot+ [
                    error bars/.cd,
                    y dir=both,
                    y explicit,
                ] 
                coordinates {(Oracle,4.54) +- (0,0.057)
                              (Propose,2.92) +- (0,0.23)
        		               (AutoVC,2.50) +- (0,0.24)
        		               (Chou et al.,2.33) +- (0,0.261)
        		               }; 
        \addplot+ [
                    error bars/.cd,
                    y dir=both,
                    y explicit,
                ] 
                coordinates {(Oracle,4.54) +- (0,0.057)
                              (Propose,3.2) +- (0,0.227)
        		               (AutoVC,2.55) +- (0,0.21)
        		               (Chou et al.,2.07) +- (0,0.261)
        		               }; 	
        \addplot+ [
                    error bars/.cd,
                    y dir=both,
                    y explicit,
                ] 
                coordinates {(Oracle,4.54) +- (0,0.057)
                              (Propose,3.25) +- (0,0.22)
        		               (AutoVC,2.48) +- (0,0.22)
        		               (Chou et al.,2.04) +- (0,0.22)
        		               };        
        \addplot+ [
                    error bars/.cd,
                    y dir=both,
                    y explicit,
                ] 
                coordinates {(Oracle,4.54) +- (0,0.057)
                              (Propose,3.46) +- (0,0.20)
        		               (AutoVC,2.88) +- (0,0.23)
        		               (Chou et al.,2.50) +- (0,0.24)
        		               };           		               
        
        \end{axis}
    \end{tikzpicture}
    \end{minipage}
    \begin{minipage}[b]{\linewidth}
        \centering
        \begin{tikzpicture}
            \begin{axis}
            [
                ybar,
                enlargelimits=0.3,
                legend style={at={(0.5,-0.35)},
                              anchor=north,legend columns=-1,font=\fontsize{8}{8}\selectfont},
                ylabel={Similarity},
                symbolic x coords={AutoVC,Chou et al.,Propose},
                xtick=data,
                nodes near coords,
                node near coords style={font=\tiny, rotate=45},
                width=\linewidth, height=3.6cm,
                ymin=1, ymax=5,
                ytick={1,2,3,4,5},
                bar width=7pt,
                ymajorgrids=true,
                grid style=dashed
                ]

            \addplot+ [
                        error bars/.cd,
                        y dir=both,
                        y explicit,
                    ] 
                    coordinates {
                                  (Propose,3.68) +- (0,0.22)
            		               (AutoVC,3.20) +- (0,0.26)
            		               (Chou et al.,2.68) +- (0,0.23)
            		               }; 
            \addplot+ [
                        error bars/.cd,
                        y dir=both,
                        y explicit,
                    ] 
                    coordinates {
                                  (Propose,3.56) +- (0,0.22)
            		               (AutoVC,3.05) +- (0,0.24)
            		               (Chou et al.,2.38) +- (0,0.22)
            		               }; 	
            \addplot+ [
                        error bars/.cd,
                        y dir=both,
                        y explicit,
                    ] 
                    coordinates {
                                  (Propose,3.62) +- (0,0.25)
            		               (AutoVC,3.09) +- (0,0.24)
            		               (Chou et al.,2.38) +- (0,0.20)
            		               };        
            \addplot+ [
                        error bars/.cd,
                        y dir=both,
                        y explicit,
                    ] 
                    coordinates {
                                  (Propose,3.72) +- (0,0.3)
            		               (AutoVC,3.04) +- (0,0.25)
            		               (Chou et al.,2.86) +- (0,0.28)
            		               };           		               
            \legend{M-M,M-F,F-M,F-F}
            
            \end{axis}
        \end{tikzpicture}
    \end{minipage}
    
\caption{MOS test scores on unseen speaker set}
\label{mos-unseen}
\end{figure}
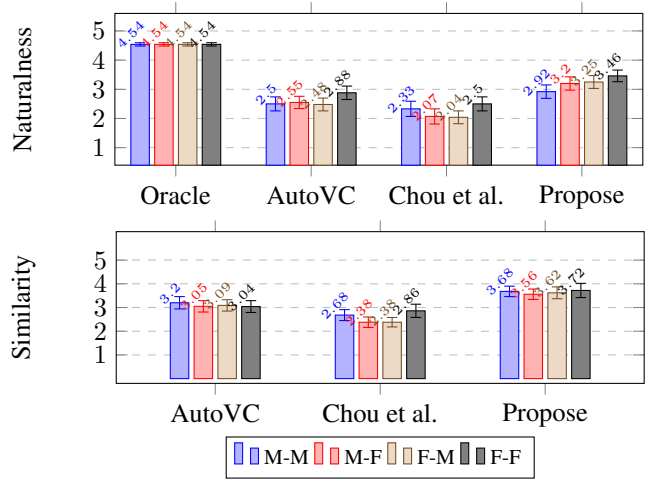

\section{Conclusion}
\label{sec:conclusion}
\vspace{-3pt}
In this paper, we proposed a novel non-parallel voice conversion model based on the disentanglement framework.
By introducing the training method for matching conversion and training procedure, the proposed model can be trained on non-parallel data more efficiently.
We designed a speaker consistency loss to directly address the conversion task on training phase.
Also, we modified the content bottleneck extraction method, and it results in the improvement of the disentanglement with little missing linguistic information.
Both objective and subjective evaluation results confirmed that the proposed method outperformed baseline methods.

\clearpage

\bibliographystyle{IEEEbib}
\bibliography{Template}

\end{document}